\documentclass[12pt]{article}
\usepackage{epsfig}
\usepackage{amssymb,amsmath,amsfonts}
\begin{document}

\title{{\bf \Large Statistics of Fluctuating Colloidal Fluid-Fluid Interfaces}}

\author{V.W.A. de Villeneuve$^{1}$, J.M.J. van Leeuwen$^{2}$,\\ 
W. van Saarloos$^{2}$ and H.N.W. Lekkerkerker$^{1}$.\\*[2mm]
$^{1}$ Van 't Hoff Laboratory for Physical and Colloid Chemistry, \\
University of Utrecht, Padualaan 8, 3584 CH Utrecht,  The Netherlands\\
$^{2}$ Instituut-Lorentz, Leiden University, Niels Bohrweg 2, \\ 
Leiden, 2333 CA, The Netherlands}
\maketitle
\noindent
PACS {\bf 05.40.-a} --Fluctuation phenomena, random processes, \\
noise and Brownian motion\\
PACS {\bf 68.03.Kn} --Fluid interfaces\\
PACS {\bf 82.70.Dd} --Colloids\\
PACS {\bf 87.64.Tt} --Confocal Microscopy\\
\begin{abstract}
Fluctuations of the interface between coexisting colloidal fluid phases have 
been measured with confocal microscopy. Due to a very low surface tension, the 
thermal motions of the interface are so slow, that a record can be made 
of the positions of the interface. The theory of the interfacial height 
fluctuations is developed. For a host of correlation functions, the 
experimental data are compared with the theoretical expressions. 
The agreement between theory and experiment is remarkably good.
\end{abstract}

\section{Introduction}\label{intro}

The study of interfaces has a long and interesting history. In 1894 van der
Waals \cite{waals} proposed an interface theory, which leads to a flat interface
with a density profile in the direction of gravity. This result is sometimes 
referred to as the {\it intrinsic interface}. Already von
Schmoluchowski \cite{smoluchowski} realized that the thermal motion of the
molecules induces height fluctuations in the interface. These motions have
been called {\it capillary waves}, since they derive from an interplay of 
gravity and surface tension, like capillary rise. The fluctuations
were first treated theoretically and experimentally 
by Mandelstam \cite{mandelstam}. He pointed out that the interface 
width diverges due to the short wavelength capillary waves.
This fact was rediscovered by Buff, Lovett and
Stillinger \cite{buff} fifty years later, after which it obtained a prominent
place in the discussion of interfaces. Weeks \cite{weeks} later pointed out 
that the notion of capillary waves only applies to wave lengths larger than
the fluid correlation length, which is of the order of the interparticle distance. 

The experimental study of interfaces
was undertaken by Raman \cite{raman} and Vrij \cite{vrij} with light scattering 
and starting with Braslau et al. \cite{braslau} by X-ray scattering. 
Although scattering on interfaces is most valuable, it always yields 
{\it global} information on the fluctuations, while a photographic inspection
gives {\it local} information.
However, the wave lengths and heights involved in the capillary waves of 
molecular fluids are way out of the reach of detection by photographic methods. 
The visual inspection of capillary waves initially remained restricted to 
computer simulations of interfaces in molecular systems \cite{sikkenk}.

The field obtained another dimension by recent experiments of Aarts et al. 
\cite{aarts} in which they obtained pictures of fluctuating colloidal 
interfaces. The key is that, by lowering the surface tension to the $nN/m$ 
range, the characteristic length and time scale of the fluctuations become 
accessible by confocal microscopy. This opened up the possibility to follow in
detail the motion of the height of the interface and to do a statistical 
analysis of its temporal and spatial behavior. Of course the method has its 
inherent restrictions. Just as in ordinary movie recording, the pixels have a
finite distance and the snapshots have to be taken 
at finite time intervals. For colloidal interfaces this interval can be made 
much smaller than the intrinsic time scale of the motions. 
Thus the Brownian character of the motion could be demonstrated {\it ad oculos}.

\begin{figure}
\begin{center}
    \centering
    \includegraphics[width=12 cm]{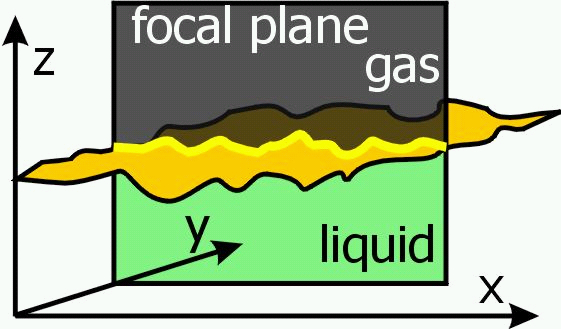}

    \caption{Schematic view of Confocal Microscopy. The confocal microscope 
thin focal planes of approximately 500 nm thickness can be imaged. This enables
the investigation of local phenomena such as height fluctuations.
}  
\label{picaarts} 
\end{center}
\end{figure}
In the confocal microscopy a two-dimensional section is inspected perpendicular
to the interface and the density profile between the two phases is observed. 
A schematic picture of the experiment is shown in Fig. \ref{picaarts}. 
A very precise location of the interface can be obtained by fitting the 
intensity with a van der Waals-like profile: $I(z,x) = a + b \tanh ([z-h(x)]/c)$, 
where $z$ is the direction perpendicular to the interface and $x$ a 
coordinate along the interface. In the upper phase the density approaches 
a value corresponding to $a+b$ and in the lower phase to $a-b$, while $c$ 
measures the intrinsic width of the interface.
Thus at every snapshot a function $h(x)$ follows and the sequence of snapshots
gives the function $h(x,t)$. This is a practical separation of the particle 
motions, which lead at short scales to the {\it intrinsic interface} and the 
particle motions which drive the long wavelength {\it capillary waves}. 
One might think that, if the time interval
of the snapshots is sufficiently small with respect to the characteristic 
time scale of variation in $h(x,t)$, one can analyze $h(x,t)$ as a continuous 
function of the time, like a movie gives the impression of a continuous
motions, while it is a succession of snapshots. In a previous short report
 \cite{villeneuve} on these experiments we have pointed 
out that the statistics remains dependent on the time interval, due to the
Brownian character of the motion. 

We have performed confocal microscopy measurements on phase separated 
colloid-polymer mixtures. The colloids are 69 nm radius fluorescently-labeled 
polymethylmetacrylate particles, suspended in cis/trans decalin, with polystyrene 
(estimated radius of gyration = 42 nm \cite{vincent}) added as depletant polymer. 
Due to a depletion induced attraction these mixtures phase separate at 
sufficiently high colloid and polymer volume fractions  and a proper colloid to 
polymer aspect ratio into a colloid-rich/polymer-poor (colloidal liquid) and a 
colloid-poor/polymer-rich (colloidal gas) phase \cite{lekkerkerker}. Here the 
polymer concentration acts as an inverse temperature.  By diluting several phase 
separating samples with its solvent decalin, the phase diagram was constructed. 
With a Nikon E400 microscope equipped with a Nikon C1 confocal scanhead, series 
of 10 000 snapshots of the interface were recorded at constant intervals 
$\Delta t$ of 0.45 s and 0.50 s of two statepoints to be denoted as II and IV. 
The pixels are separated by a distance $\Delta x = 156nm$ and a single scan 
takes approximately 0.25 s to complete.

The set-up of the paper is as follows. We start out by discussing the spatial 
behavior of the data of a single time frame, which requires only equilibrium 
statistics. The correlation functions and the statistics of hills and valleys 
in the interface are determined and compared to the theory.

Then we identify the set of interface modes via the fourier decomposition 
\begin{equation} \label{a1}
h({\bf x},t) = \sum_{\bf k} h_{\bf k} (t) \exp(i{\bf k \cdot x}) .
\end{equation} 
The modes are overdamped in the relevant regime and follow from the 
macroscopic interface-dynamics. The motion obeys not only the macroscopic 
equations but is also influenced by noise. We introduce thermal noise through 
the Langevin equation and calculate the essential height-height correlation 
function $\langle h({\bf 0},0)h({\bf x},t) \rangle$. Via the equivalent 
Fokker-Planck equation the probabilities on sequences (``histories'') of 
snapshots are determined.  The analysis of the distributions of ``hills'' and 
``valleys'' in the time domain with respect to a level $h$ is similar to the spatial 
behavior. A special concern is the dependence of the residence time 
and the waiting time on the used time-interval. 

The paper closes with a discussion of the main results. 
 
\section{Equal time Correlations} \label{equal}

The function $h({\bf x},t)$ provides a mathematical division between the two 
coexisting phases which form the interface. The interface is 
of the solid-on-solid type since so-called overhangs, well-known in lattice
theory, are excluded by construction, as to every value of the horizontal
coordinate ${\bf x}$ and time $t$ one unique height $h({\bf x},t)$ is associated.
The basic function is the height-height correlation function.
Due to translational invariance the modes ${\bf k}$ are independent and thus the 
correlation function in space has the fourier decomposition
\begin{equation} \label{c0}
\langle h({\bf 0}, 0) h({\bf x}, 0) \rangle = 
\sum_{\bf k} \langle |h^2_{\bf k}| \rangle \exp(i{\bf k \cdot x})
\end{equation} 
The brackets denote equilibrium averages and $h_{\bf k}$ is the amplitude of the
${\bf k}$-th mode. The distribution of the $h_{\bf k}$ follows from the Boltzmann
factor involving the energy of a deformation of the interface, which is given by 
the drumhead model
\begin{equation} \label{c4}
{\cal H} (\{h\}) = {1 \over 2} \int  d{\bf x} \left[ \Delta \rho g h^2({\bf x}) + 
\gamma (\nabla h({\bf x}))^2 \right].
\end{equation}
Here $\Delta \rho$ is the density difference between the coexisting phases, 
$\gamma$ is the surface tension and $g$ is the gravitational acceleration. 
The first term gives the gravitational potential energy and the second term the
increase of the interfacial energy due to curvature.  Expressed in terms of the
amplitudes $h_{\bf k}$ it reads
\begin{equation} \label{c5}
{\cal H} (\{h\}) = {L^2 \over 2}\sum_{\bf k} [\Delta \rho g  + \gamma k^2]\, 
|h_{\bf k}|^2,
\end{equation} 
where $L^2$ is the area of the interface. Since (\ref{c5}) is quadratic in the 
amplitudes $h_{\bf k}$, it implies a gaussian distribution for the $h_{\bf k}$ 
\begin{equation} \label{c6}
P_e(h_{\bf k}) = { \exp -|h_{\bf k}|^2/ 2 \langle |h_{\bf k}|^2  \rangle \over 
[2 \pi \langle |h_{\bf k}|^2 \rangle]^{1/ 2}},
\end{equation} 
with the average 
\begin{equation} \label{c7}
\langle |h_{\bf k}|^2  \rangle = {k_B T \over L^2 (\Delta \rho g + \gamma k^2)}.
\end{equation} 
\begin{figure}
\begin{center}
    \centering
    \includegraphics[width=12 cm]{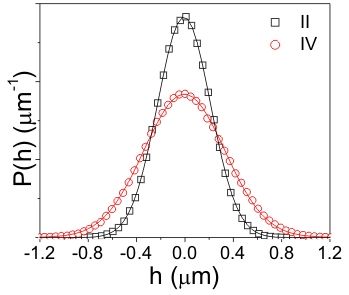}
    \caption{The height distribution for statepoint II and IV as found 
experimentally. The lines are gaussian fits to the data.}
\label{heights} 
\end{center}
\end{figure}

With the distribution (\ref{c6}) of the $h_{\bf k}$, we can calculate the 
distribution of the heights $h$, which becomes also a gaussian
\begin{equation} \label{c8}
P_{eq} (h) = {\exp (-h^2 /2 \langle h^2 \rangle) \over [2 \pi 
\langle h^2 \rangle]^{1/ 2}},
\end{equation} 
with the mean square height $\langle h^2 \rangle$ given by
\begin{equation} \label{c9}
\langle h^2 \rangle = \sum_{\bf k} \langle h^2_{\bf k}  \rangle =
{k_B T \over 4 \pi \gamma} \ln {1+ k^2_{max} \xi^2 \over 1+ k^2_{min} \xi^2}.
\end{equation}
$\xi$ is the capillary length defined as
\begin{equation} \label{x0}
\xi^2 = {\gamma \over g \Delta \rho}.
\end{equation}  
The integral has been given an upper bound $k_{max} \simeq 2 \pi/d$
with $d$ the diameter of the particles and a lower bound 
$k_{min} \simeq 2 \pi/L$ due to the finite size of the interface.
The lower bound can be set equal to 0 for all practical purposes, but the
upper bound is essential for the convergence of the integral. 
Cutting off the capillary waves at the short-wavelength side is the poor man's
way to handle the otherwise diverging interface width $\langle h^2 \rangle$. 
There are two options to determine 
$\langle h^2 \rangle$. The first follows from a fit to $P_{eq} (h)$, 
which is shown in Fig. \ref{heights}. The second is a direct evaluation
of $\langle h^2 \rangle$ from the recorded data.
The latter always gives a 1-3 \% larger value, which we attribute to 
optical artifacts due to confocal slicing. So we are inclined to 
prefer the former value which amounts to $\langle h^2 \rangle= 0.219 $
for statepoint II and 0.336 $(\mu m)^2$ for statepoint IV. Then equation 
(\ref{c9}) can be used to estimate the upper cut-off. On the basis of a 
determination of $\gamma$ (see below) one finds values 
around $\kappa = k_{max} \xi \simeq 45$, but this value is rather sensitive 
to small variations in $\gamma$: a variation of $\gamma$ by 10-15\% results in a shift in $\kappa$ by a factor 2.

\begin{figure}
\begin{center}
    \centering
    \includegraphics[width=12 cm]{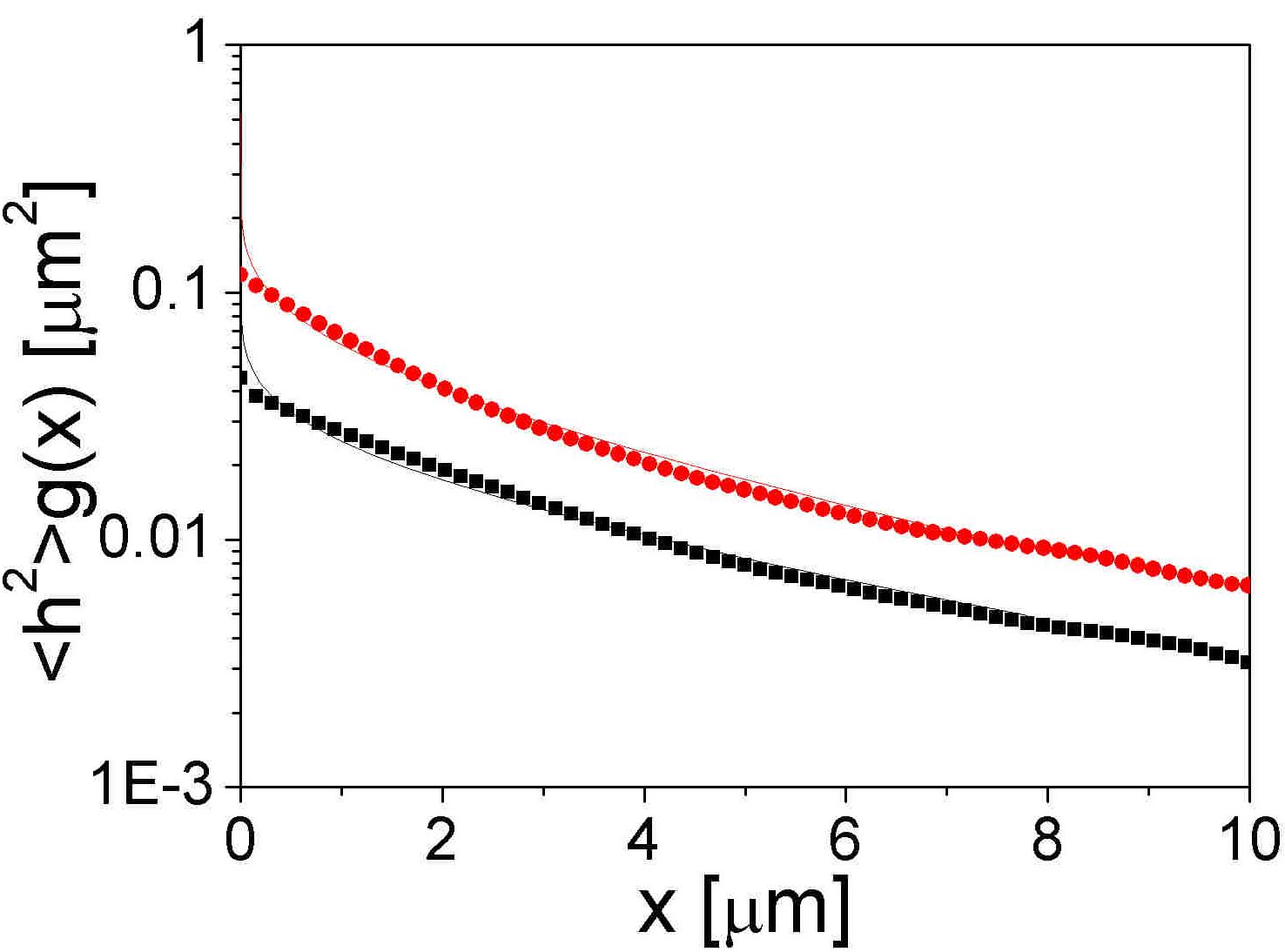}
    \caption{The spatial correlation function $\langle h(x,0)h(0,0\rangle$
for the statepoints II and IV fitted to the expression (\ref{cx}) using a 
cut-off. The lines are fits to the experimental data.}
\label{gxfits} 
\end{center}
\end{figure}

The correlation function $\langle h({\bf 0}, 0) h({\bf x}, 0) \rangle$ is 
of course also measurable. In the appendix we discuss the integral (\ref{c0}); 
here we give the result with the cut-off sent to $\infty$ 
\begin{equation} \label{cx}
\langle h({\bf 0},0) h({\bf x},0)\rangle = {k_B T \over 2 \pi \gamma} K_0(x/\xi).  
\end{equation} 
The divergence for $x \rightarrow 0$ of the modified Bessel function $K_0$ 
corresponds to the divergence of the interface width without a cut-off.
A fit of the correlation function to the Besselfunction (with a slight 
modification due to the cut-off) is shown in Fig. \ref{gxfits}. Apart from
a few initial points the function fits quite well. We find fitting parameters 
$\xi = 8.0 \mu m$ for statepoint II and $\xi=6.1 \mu m$ for statepoint IV.
The values for $\gamma$ turn out to be $ 58 nN/m$ viz $21 nN/m$ for statepoint
II viz. IV. 

Fourier transforming the correlation function back to the wavenumber domain
should lead to the expression (\ref{c7}) as function of $k$. However, an inverse
fourier transform requires accurate data for a large domain and the correlation
function is unreliable for large distances (not shown in Fig. \ref{gxfits}).
This prevents a direct check of the drumhead hamiltonian. 

\section{Multiple Correlation Functions}

As the data are stored for all sampled positions we can determine the 
probability density

\begin{multline} \label{x1}
G_n (h_1,{\bf x}_1; \cdots ; h_n,{\bf x}_n) = \\
\langle \delta (h({\bf x}_1,0)
 - h_1) \cdots \delta (h({\bf x}_n,0) - h_n) \rangle,
\end{multline} 
which gives the joint probability that the interface at position ${\bf x}_1$ has 
the height $h_1$ and subsequently at position ${\bf x}_i$ the height $h_i$ etc..
A straightforward evaluation of (\ref{x1}) proceeds via writing the $\delta$ 
functions as a fourier integral and then expressing $h({\bf x},0)$ in terms of
the amplitudes $h_{\bf k}$. As all  integrals are over a quadratic form in
the exponent the result is a gaussian in the $h_i$. So one gets
an the expression of the type
\begin{multline} \label{x2}
G_n (h_1, {\bf x}_1; \cdots ; h_n, {\bf x}_n) = \left({\det g^{-1} \over (2 \pi \langle h^2 \rangle)^n }
\right)^{1/2} \\
 \exp \left[ - {1\over 2} \sum_{i,j} g^{-1}_{i,j} {h_i h_j \over \langle 
h^2 \rangle} \right].
\end{multline} 
In this notation the matrix $g_{i,j}$ is the correlation matrix
\begin{equation} \label{d5}
g_{i,j} = g(|{\bf x}_i-{\bf x}_j|, 0),
\end{equation} 
which turns out to be the equal-time height-height correlation function
\begin{equation} \label{x3}
g(|{\bf x}-{\bf x}'|, t-t')=  \langle h({\bf x}, t) h({\bf x}', t') \rangle / 
\langle h^2 \rangle.
\end{equation}
A shortcut to the result (\ref{x2}) is to evaluate the following integral in two 
ways
\begin{multline} \label{d6}
\int dh_1 \cdots dh_n \, h_i h_j \, G(h_1,{\bf x}_1; \cdots ; h_n, {\bf x}_n) = \\
\langle h^2 \rangle g_{i,j}.
\end{multline}
The first uses the definition (\ref{x1}) and obviously leads to the right hand 
side of (\ref{d6}). The second way uses the expression (\ref{x2}). Then one has
to diagonalize the quadratic form in the exponent and the integration over
the eigendirections also leads to the right hand side of (\ref{x2}), which shows
that (\ref{d5}) is correct.

In (\ref{x3}) and accordingly in the result (\ref{x2}), we have factored out 
$\langle h^2 \rangle$ because we want to use it as a scale for the heights.
(\ref{x2}) shows that the correlation function $g$ dictates the 
behavior of the multiple correlation functions.

To give an impression on the behavior of the $G$'s we first consider a few small
values of $n$. For a single position ($n=1$) the value $g(0,0)=1$ and 
(\ref{x2}) reduces to the equilibrium height distribution (\ref{c6})
\begin{equation} \label{e1}
G_1 (h_1) = P_{eq} (h_1),
\end{equation} 
which is shown in Fig. \ref{heights}. 
From this probability we derive the important probabilities $q^+ (h)$ to 
find a height above $h$ and $q^- (h)$ for finding a height below $h$. 
They are given by the expressions
\begin{multline} \label{e2}
q^+ (h) = \int^\infty_h P_{eq} (h_1) dh_1, \\
q^- (h) = \int^h_{-\infty} P_{eq} (h_1) dh_1.
\end{multline}
In integrals like (\ref{e2}), one changes of course to the combination 
$h_1 / \langle h^2 \rangle^{1/2}$ as integration variable, such that the 
$q^\pm(h)$ become functions of the scaled variable 
$h / \langle h^2 \rangle^{1/2}$. The result of the integration in (\ref{e2}) is
an error function in this parameter. From now on we work with these 
reduced heights. 

The probability density $G$ for $n=2$ is still sufficiently simple to make it 
explicit. The matrix $g_{i,j}$ and its inverse $g^{-1}_{i,j}$ read 
\begin{equation} \label{e3a}
g_{i,j} =\left(
\begin{array}{cc}
1 & g_{1,2} \\*[2mm]
g_{2,1} & 1 \\*[2mm]
\end{array} \right),
\end{equation}
and
\begin{equation}\label{e3b}
g^{-1}_{i,j} ={1 \over 1- g^2_{1,2}} \left(
\begin{array}{cc}
1 & -g_{1,2} \\*[2mm]
- g_{2,1} & 1 \\*[2mm]
\end{array} \right). 
\end{equation} 
So $G_2$ follows  from the general definition as
 \begin{multline} \label{e4}
G_2 (h_1,{\bf 0}; h_2,{\bf x}) = {1 \over 2 \pi  [1-g^2]^{1/2}}
\exp \\ \left[ - {h^2_1 - 2 g h_1 h_2 + h^2_2 \over 2 [1 - g^2] } \right],
\end{multline} 
with $g$ a shorthand for $g_{1,2} = g (x,0)$ and $x$ the distance of sampling.
Note that this expression is symmetric in the entries $h_1$ and $h_2$ and that 
dependence only enters through $g=g(x,0)$. 

\section{Statistics of Sequences} \label{sequence}

The probability densities (\ref{x1}) are measurable, but the statistics becomes 
poor when too much entries are taken. Therefore integrated probabilities
are more accessible. For what follows it is interesting to study the probability
that a sequence in space of precisely $n$ successive values occurs of the heights
{\it above} the level $h$. According to the theory it is given by the ratio of 
two integrals
\begin{equation} \label{e8}
p^+_n (-h) = q^{-(n+)-} (h) / q^{+-} (h).
\end{equation} 
In this notation the superscript prescribes the integration domain. 
The numerator of (\ref{e8}) reads
\begin{multline} \label{e10}
 q^{-(n+)-} (h)  = \int^h_{-\infty} dh_0 \int^\infty_h dh_1 \cdots \\
\int^\infty_h dh_n
\int^h_{-\infty} dh_{n+1}  G_{n+2}  (h_0, \cdots, h_{n+1}).
\end{multline} 

\begin{figure}
\begin{center}
    \centering
    \includegraphics[width=12 cm]{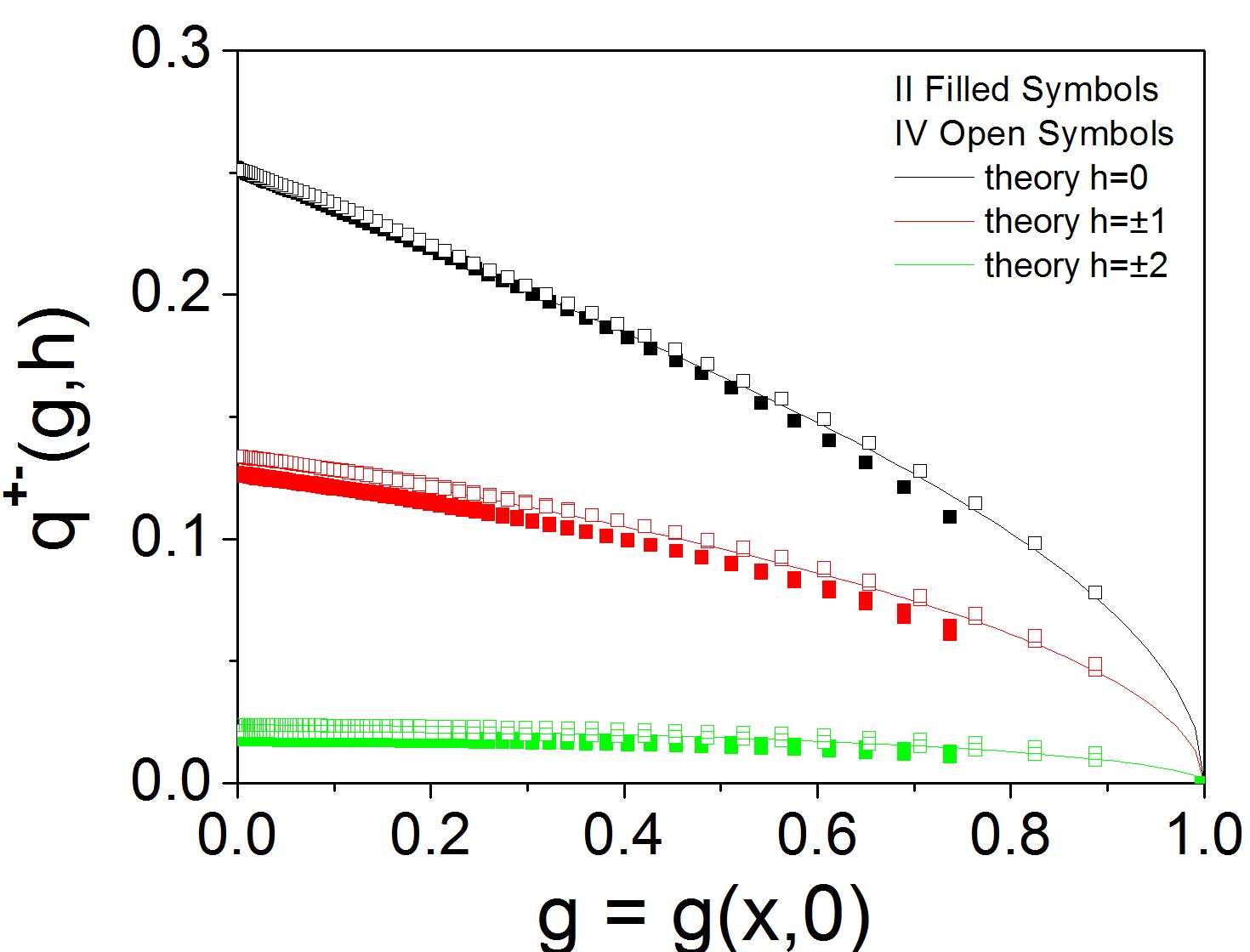}
    \caption{The function $q^{+-} (h,g) $ as function of $x$ through
$g=g(x,0)$, for some values of $h$ (in units $\langle h^2 \rangle^{1/2})$.
The drawn lines are calculated and the points are the measured values.}
\label{mountx} 
\end{center}
\end{figure}

The integral over the first variable $h_0$ guarantees that the sequence
starts below level $h$, the next $n$ integrations select points above the 
level $h$ and the sequence ends with $h_{n+1}$ below level $h$. So the numerator
in (\ref{e8}) selects the hills of precisely $n$ consecutive values of the
height above level $h$. We have omitted in $G_{n+2}$ the position arguments
since it is understood that points are equidistant. So the sequence of values
$g(m \Delta x,0)$ enters, with $0 \leq m < n+2$.
The denominator is the integral
\begin{equation} \label{e9}
 q^{+-} (h)  = \int^\infty_h dh_1
\int^h_{-\infty} dh_2  G_2  (h_1, h_2).
\end{equation} 
It counts the number of hills since each hill is followed by a transition from
above to below the level $h$. The denominator serves as a normalizing factor.
Summing (\ref{e8}) over $n$ (from 1 to $\infty$) gives the total number of hills
above $h$ and as this equals the number of crossings, we see that the 
distribution (\ref{e8}) is normalized. 

The average length $\chi^+ (h)$ of a sequence is defined as
\begin{equation} \label{x4}
\chi^+ (h) = \sum_{n=1} n p^+_n (h).
\end{equation} 
We also look to sequences {\it below} the height $h$. They are  given by the 
probability $p^-_n (h)$ which follows from a similar definition as (\ref{e8}), 
with $+$ and $-$ interchanged. The up-down symmetry of the problem yields the 
relation
\begin{equation} \label{e6}
p^-_n (h) = p^+_n (-h).
\end{equation}
The average length $\chi^- (h)$ of a stretch below $h$ similarly equals
\begin{equation} \label{x5}
\chi^- (h) = \sum_{n=1} n p^-_n (h).
\end{equation} 
Inserting the expression (\ref{e8}) into the definition (\ref{x4}) for 
$\chi^+ (h)$, the numerator in (\ref{e10}) is multiplied by the 
number of values larger than $h$. Summation over all $n$ leads to the 
average number of points above the level $h$, which is given 
by the integral (\ref{e2}). Thus we arrive at the relations
\begin{equation} \label{f2}
\chi^\pm (h) = q^\pm (h) / q^{+-} (h).
\end{equation} 

The remarkable point about these relations is that, although the probabilities
$p^\pm_n (h)$ are given by multiple integrals, the averages $\chi^\pm (h)$
result from simple integrals. The $q^\pm (h)$ are errorfunctions and $q^{+-}$
is a two-fold integral involving the function $G_2$, thus containing only 
the value $g(\Delta x,0)$.

\begin{figure}
\begin{center}
    \centering
    \includegraphics[width=12 cm]{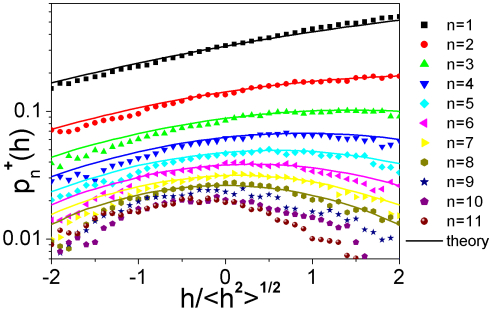}
    \caption{The spatial $p^+_n (h)$ as function of $h$ for a number of $n$ for statepoint 
IV. The drawn lines are the theoretical values.}
\label{pnplu} 
\end{center}
\end{figure}

A trivial result from (\ref{f2}) is that the ratio $\chi^+ (h)/\chi^- (h)$ is
the same as the ratio $q^+ (h)/q^- (h)$. Both give the ratio of the total number
of points above and below the level $h$. As the $q^+ (h)$ and $q^- (h)$ add 
up to 1, a more intriguing result follows for the sum
\begin{equation} \label{f6}
\chi^+ (h) + \chi^- (h) ={ 1 \over q^{+-} (h,g)}.
\end{equation} 
Deliberately we have given $q^{+-} $ also the argument $g$ which incorporates
the spatial dependence on $x$.  Tacitly we have assumed that this distance
is the sampling distance $\Delta x$. But nothing prevents us from taking a 
multiple $n$ of $\Delta x$. Then $g$ will refer to $g(n \Delta x,0)$. In Fig. \ref{mountx}
we have plotted the experimental values of $q^{+-} (h,g)$ for various values of
$g(n \Delta x,0)$, which we take as parameter on the horizontal axis. The curves
are the calculated values of $q^{+-} (h,g)$. We have not 
found a closed expression for $q^{+-}$ in terms of known functions, 
but a number of limits are explicitly obtainable. The $g$ dependence is 
exemplified by the case $h=0$, which reads
\begin{equation} \label{f3}
q^{+-} (0,g) = {1 \over 2} - {1 \over \pi} \arctan 
\left({1+g \over 1-g}\right)^{1/2}.
\end{equation}
The $h$ dependence is by and large controlled by the limiting behavior 
\begin{multline} \label{f4}
q^{+-} (h,0) = q^+ (h) q^- (h), 
\\
 q^{+-} (h, g \rightarrow 1) \simeq
{\sqrt{1-g} \over \pi \sqrt{2}} \exp \left[ - {h^2 \over 2 \langle h^2 \rangle}
\right].
\end{multline} 

Apart from the averages also the individual $p^\pm_n (h)$ can be measured and
compared with the theoretical expressions (\ref{e8}). In Fig. \ref{pnplu}
we show $p^+_n (h)$ for statepoint IV as a function of $h$ for a number of $n$. 
The theory requires the evaluation of multiple integrals (\ref{e10}) which can 
be carried out by Monte-Carlo integration. The best procedure is to generate a 
distribution according to the gaussian integrand and then reject the points 
that fall outside the integration domain. This technique also applies to 
correlation functions for other histories with another integration domain. 

The comparison with the theory has been carried out for $n$ up to 8. For higher
values of $n$ the integration becomes a bit lengthy for good statistics. The 
agreement between theory and experiment is good, but typically there are deviations for larger negative values of $h$, where the experimental points are systematically lower than the theorical prediction.

\section{The Dynamic Interface Modes} \label{modes}

As the preceding sections show, the interface fluctuations have a rich spatial 
structure. So it is an interesting question how this compares with the 
interface fluctuations in time. 
The temporal development of the interface is determined by the macroscopic
equations for the interface modes as well as by the influence of thermal noise.
In this section we briefly discuss the interface modes and in the next section 
we treat the noise.

The problem of the interface modes modes has been addressed by Jeng et al. 
\cite{jeng}, who have made an extensive study of the interface modes in the 
various regimes distinguished by the relative strength of viscosity and 
surface tension.  The modes are overdamped for our experimental conditions  
and decay as
\begin{equation} \label{b1}
h_{\bf k} (t) = h_{\bf k} \exp(-\omega_k t),
\end{equation} 
with a rate
\begin{equation} \label{b2}
\omega_k = {1 \over 2 t_c} [(k \xi)^{-1} + k \xi],
\end{equation}
where the capillary time $t_c$ is given by
\begin{equation} \label{b3} 
t_c = {(\eta+\eta') \over \sqrt{g \gamma\Delta \rho}}.                   
\end{equation}
Here $\eta$ and $\eta'$ are the viscosities of the lower and upper fluid.  
A few remarks on (\ref{b1}) are worth making:
\begin{itemize}
\item The dispersion relation $\omega_k$ is in general rather complicated. The 
simplification (\ref{b2}) derives from the approximation $\rho \omega_k \ll
\eta k^2$ which is very well fulfilled for colloidal interfaces with extremely
low surface tension. The approximation is controlled by the number 
\begin{equation} \label{b4}
{\xi \over L_\eta} = \left({\gamma^3 \Delta \rho \over g }\right)^{1/2} 
{1 \over (\eta + \eta')^2 },
\end{equation} 
which is the ratio of the capillary length $\xi$ to the viscous length 
$L_\eta= (\eta + \eta')^2 /\gamma \Delta \rho$.
It is very small, $10^{-5}$,  for colloidal interfaces, while
it is very large for e.g. water ($10^5$).
\item The spectrum has a slowest mode with wavelength $\xi$ and decay rate 
$t_c$, in contrast to the capillary waves of molecular fluids, where the 
modes become slower the longer the wavelength. This mode starts to dominate 
the behavior of the correlations for long times.
\end{itemize}

\section{The Langevin Equation}\label{langevin}

The thermal influences can be incorporated by a 
fluctuating force $F_{\bf k} (t)$ on mode ${\bf k}$ in the Langevin equation 
\cite{kampen}
\begin{equation} \label{c1}
{\partial h_{\bf k} \over \partial t} = - \omega_k h_{\bf k} + F_{\bf k} (t).
\end{equation} 
The first term on the right hand side is the systematic damping force, which by 
itself would lead to an exponential decay of mode $k$. The random force
$F_{\bf k}(t)$ has zero average and 
is assumed to be $\delta$--correlated in time (white noise)
\begin{equation} \label{c2}
\langle F_{\bf k} (t)\, F_{\bf k'} (t') \rangle = \delta_{\bf k+k',0} \, \Gamma_{\bf k} 
\, \delta(t-t'),
\end{equation}
where $\Gamma_{\bf k}$ can be found from the fluctuation-dissipation theorem
\begin{equation} \label{c3}
{\Gamma_{\bf k} \over 2 \omega_k} = \langle |h_{\bf k}|^2 \rangle. 
\end{equation}
The Langevin equation (\ref{c1}) assumes that the slow capillary waves form
a complete set to characterize the motion of the interface. 
$\Gamma_{\bf k}$ is linked in (\ref{c3}) to the equilibrium average of the 
amplitudes $h_{\bf k}$, which is given by (\ref{c6}).

With the Langevin equation all time-dependent correlation functions can 
be calculated.  In particular the height-height correlation function follows as
\begin{multline} \label{c10}
\langle h({\bf 0},0) h({\bf x},t) \rangle = \langle h^2 \rangle g(x, t) = \\
\sum_{\bf k} \langle |h_{\bf k}|^2 \rangle \exp (i{\bf k \cdot x} - \omega_k t), 
\end{multline} 
with $ \langle |h_{\bf k}|^2 \rangle$ given by (\ref{c7}) and $\langle h^2 \rangle$
by (\ref{c9}). Note that for this correlation function the influence of the 
fluctuating force $F_{\bf k} (t)$ averages out such that it is depends only 
on the macroscopic dynamics of the interface. It involves, apart from the
decay rate $\omega_k$ only the thermal average $\langle |h_{\bf k}|^2 \rangle$.
Some properties of the integral yielding this function are listed in the Appendix.
 
The first point is the determination of $\gamma$ and $t_c$ from the data. We
represent $\langle h({\bf 0},0) h({\bf 0},t) \rangle$ as
\begin{equation} \label{c11}
\langle h({\bf 0},0) h({\bf 0},t) \rangle = {k_B T \over 2 \pi \gamma } 
H(t/t_c , \kappa),
\end{equation}
Here again $\kappa = k_{max} \xi$. In the Appendix we prove that
\begin{equation} \label{c12}
H(t/t_c, \infty) = K_0 (t/t_c),
\end{equation}
with $K_0 $ the modified Besselfunction of order 0. 
\begin{figure}
\begin{center}
    \centering
    \includegraphics[width=12 cm]{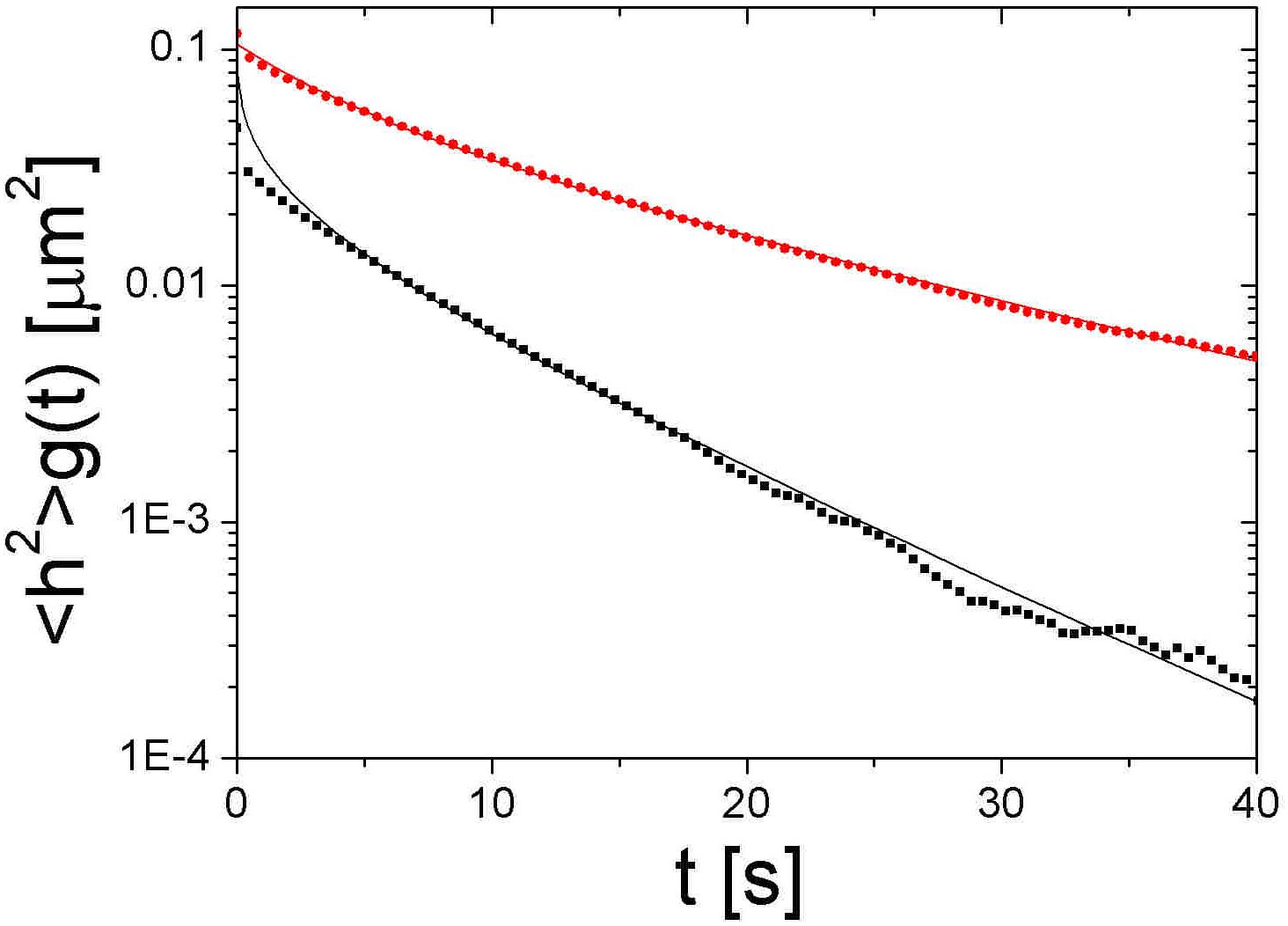}
    \caption{The correlation function $\langle h(0,t)h(0,0) \rangle$. The points are 
experimental values and the lines curves according to (\ref{c12}) using a cut-off.}
\label{correlation} 
\end{center}
\end{figure}
For $t \geq t_c / \kappa$, the function $H(t/t_c , \kappa)$ is well 
represented by $K_0 (t/t_c)$. Since $\kappa$ is of the order 40 to 50 
(see Section \ref{equal}), expression (\ref{c12}) suffices for most of  the 
measured points, except of course for the first few points near $t=0$, where 
the right hand side of (\ref{c12}) diverges. Leaving them out for the moment,
we find from a fit for statepoint II: 
$t_c = 20 s$ and $\gamma = 66 nN/m$ and for statepoint IV: $t_c = 33 s$ 
and $\gamma =22 nN/m$. Effectively $t_c$ acts as a horizontal scale
parameter and $\gamma$ as a vertical shift. $t_c$ is mainly determined by
the asymptotic behavior, while $\gamma$ is more sensitive to the initial behavior.

By adjusting the 
upper cut-off, the calculated $g(0,t)$ assumes the value 1 for $t=0$.
In Fig. \ref{correlation} we plot the experimental values of $\langle h^2 \rangle g(0,t)$ 
together with the theoretically calculated curves.

Finally we mention the initial behavior of $g(0,t)$. From the expansion (\ref{A10})
we deduce
\begin{equation} \label{c14}
g(0,t) = 1 - {t \over t_c} {\kappa \over \ln (1+ \kappa^2)} + \cdots
\end{equation} 
Here one sees that a finite $\kappa$ is essential for this initial behavior.

\section{Probabilities on Histories}\label{correl}

The noise term comes into the picture when we calculate the distribution
of the $h_{\bf k} (t)$. It follows from
the Fokker-Planck equation, which is equivalent with the Langevin equation
and reads \cite{kampen}
\begin{equation} \label{d1}
{\partial P(h_{\bf k},t) \over \partial t} = \omega_k  \,{\partial h_{\bf k} 
P(h_{\bf k},t) \over \partial h_{\bf k}} + {\Gamma_{\bf k} \over 2} \, 
{\partial^2 P(h_{\bf k},t) \over \partial h_{\bf k}^2}.
\end{equation} 
It gives the evolution of the probability distribution $P(h_{\bf k},t)$ starting 
from an initial distribution $P(h_{\bf k},0)$. The solution \cite{kampen} of 
(\ref{d1}) provides the conditional probability of the mode $h_{\bf k} (t)$, 
starting with the value $h_{\bf k} (0)$
\begin{multline} \label{d2}
P(h_{\bf k} (0)| h_{\bf k}(t)) = {1 \over [2 \pi \langle |h_{\bf k}|^2 \rangle
(1-e^{-2 \omega_k t})]^{1/2}} \\
\exp - {|h_{\bf k}(t) - h_{\bf k}(0) e^{-\omega_k t}|^2 
\over 2 \langle |h_{\bf k}|^2 \rangle(1-e^{-2 \omega_k t})}.
\end{multline}
The expression shows that, independent of the value of $h_{\bf k} (0)$, the 
distribution asymptotically approaches the equilibrium distribution (\ref{c6}).

For the measurements at different times (and possibly different positions)
we need the multiple time correlation function
\begin{multline} \label{d3}
G_n (h_1,{\bf x}_1,t_1; \cdots ; h_n,{\bf x}_n,t_n) = \\ \langle \delta (h({\bf x}_1,t_1)
 - h_1) \cdots \delta (h({\bf x}_n,t_1) - h_n) \rangle,
\end{multline}
giving the probability of a {\it history} that the interface is at time $t_1$ 
and position ${\bf x}_1$ at a height $h_1$ and subsequently at time $t_i$ and 
position ${\bf x}_i$ at height $h_i$. In order to evaluate these correlation 
functions we have to translate the field $h({\bf x},t)$ into its fourier 
components $h_{\bf k}$. Then we have to use the joint probability on a set of 
components $h_{\bf k} (t_1), \cdots h_{\bf k}(t_n)$, which is given by the product 
of the equilibrium probability (\ref{c9}) for the first event at $t_1$ and the 
conditional probabilities (\ref{d2}) for the successive time intervals, 
$t_2-t_1, \cdots, t_n-t_{n-1}$. The result of the integration can be derived easily
from the observation that the $h({\bf x}_j,t_j)$ are, as linear combinations of the
basic variables $h_{\bf k}$, also gaussian random variables. So, similar to the
derivation of  (\ref{x2}), their distribution must be of the form
 \begin{multline} \label{d4}
G_n (h_1,\cdots , t_n) = \left({\det g^{-1} \over (2 \pi \langle h^2 \rangle)^n }
\right)^{1/2} \\
\exp \left[ - {1\over 2} \sum_{i,j} g^{-1}_{i,j} {h_i h_j \over \langle 
h^2 \rangle} \right].
\end{multline} 
The matrix $g_{i,j}$ is the correlation matrix
\begin{equation} \label{d5a}
g_{i,j} = g(|{\bf x}_i-{\bf x}_j|, t_i-t_j)
\end{equation} 
The proof of (\ref{d4}) is exactly the same as that of (\ref{x2}).

(\ref{d4}) is the main result of the theory for the histories. 
It relates the probability of a 
history $h_1, \cdots, h_n$ on a sequence of snapshots to the height-height 
correlation function $g(x,t)$. The strong point of (\ref{d4}) is that the 
time-dependent probability densities have exactly the same structure
as the equal time probabilities, when expressed in the appropriate $g_{i,j}$.
Thus the whole analysis given above for the equal time correlations, can be
taken over for the more general correlations. So we restrict ourselves
for the time dependent histories to the aspects needing some extra attention. 

The time dependent probability density $G_2$ reads as (\ref{e4}) 
with $g=g(0,t)$. It can also be written as  
the product of the equilibrium distribution $P_{eq} (h_1)$ and the conditional 
probability $G_c (h_1,0,0 | h_2,0,t)$ that starting at $h_1$ one 
arrives at $h_2$ at time $t$ later
\begin{multline} \label{e5}
G_c( h_1,0,0 | h_2,0,t ) = {1 \over [2 \pi (1 -g^2)]^{1/2}}
\\ \exp - {[h_2 - h_1 g]^2 \over 2  [1-g^2]}.
\end{multline}
This expression cannot be seen as the ``propagator'' for the probability, 
like (\ref{d2}) is for the fourier components $h_{\bf k}$. 
While the probabilities for the modes ${\bf k}$ evolve as a Markov process, the 
distribution for $h({\bf 0},t)$ does have a memory effect. Only if $g(0,t)$ were a 
pure exponential the spatial process would be Markovian too \cite{kampen}. 
The expression (\ref{c10}) shows that $g(0,t)$ it is not a pure exponential but 
a superposition of exponentials. For longer times it starts to decay as an 
exponential when the slowest mode begins to dominate, as can be seen in 
Fig. \ref{correlation}.

\section{Average numbers of hills and valleys} \label{exper}

Consider now a sequence of snapshots, taken
with time intervals $\Delta t$.  We are again interested in the  probabilities on 
the duration of ``hills'' and ``valleys'' with respect to a level $h$. 
To stress the analogy between space and time we use the same  notation 
$p^\pm_n (h)$ for the probabilities
to find a stretch of exactly $n$ consecutive values of the interface height 
above/level the level $h$, where $n$ now is an index in the time direction.  
They are given by the same integrals as (\ref{e10})
and (\ref{e9}) with $g_{i,j}$ the temporal correlation matrix.
The mean values are called the {\it residence time} $\tau^+ (h)$ 
(for $p^+_n (h)$) and the {\it waiting time} $\tau^- (h)$ (for $p^-_n (h)$).
\begin{figure}
\begin{center}
    \centering
    \includegraphics[width=12 cm]{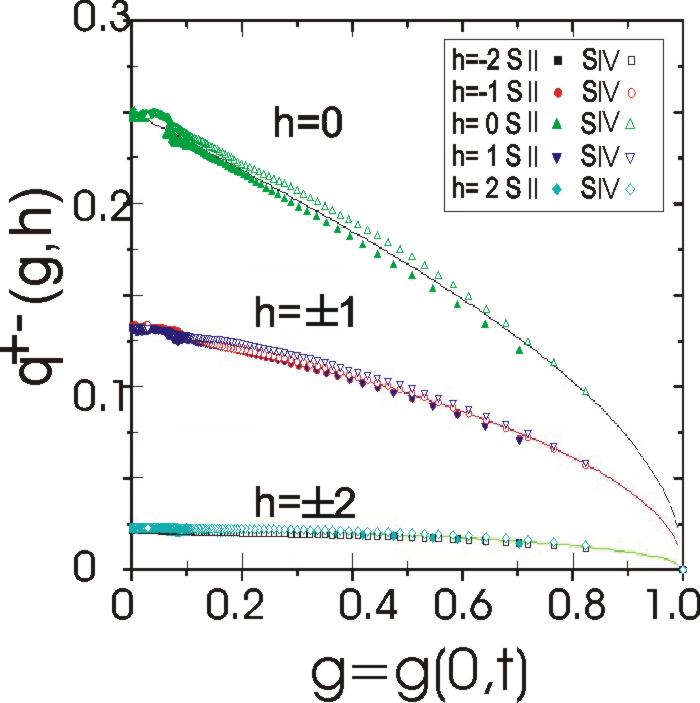}
    \caption{The function $q^{+-}$ as function of $t$ represented by 
$g=g(0,t)$, for $h=-1$ (filled
squares), $h=0$ (circles) and $h=1$ (semifilled pentagons). The colours 
correspond to time intervals $\Delta t =$ 0.5 seconds (black symbols), 
$2 \Delta t$ (red symbols) and $4 \Delta t $ (green symbols).
The drawn lines are calculated and the points are the measured values.}
\label{mounnum} 
\end{center}
\end{figure}

To check whether the experimental values follow these theoretical predictions,
we first checked that the ratio $\tau^+ (h)/\tau^- (h)$ is the same 
$q^+ (h) /q^- (h)$ in analogy with \ref{f2}. It is valid over several orders of magnitude. Only for the 
very large $h$ deviations occur due to poor statistics.
In Fig. \ref{mounnum} we now plot again the calculated values of $q^{+-} (h,g)$ 
as function of the parameter $g=g(0,t)$, for a number of $h$ values. 
The upper curve in Fig. \ref{mounnum}
for $h=0$, is given by equation (\ref{f3}). 
In this figure the experimental values are plotted as follows.
We take as time interval a multiple $n$ of the smallest interval $\Delta t$ and 
determine for this sampling rate the $\tau$'s. This leads to experimental values 
of $q^{+-} (h,g)$, which we plot in the figure at the value $g=g(n \Delta t)$. 
The curves for a fixed value of $h$ are statepoint 
independent; the figure shows that this is pretty well the case.
\begin{figure}
\begin{center}
    \centering
    \includegraphics[width=12 cm]{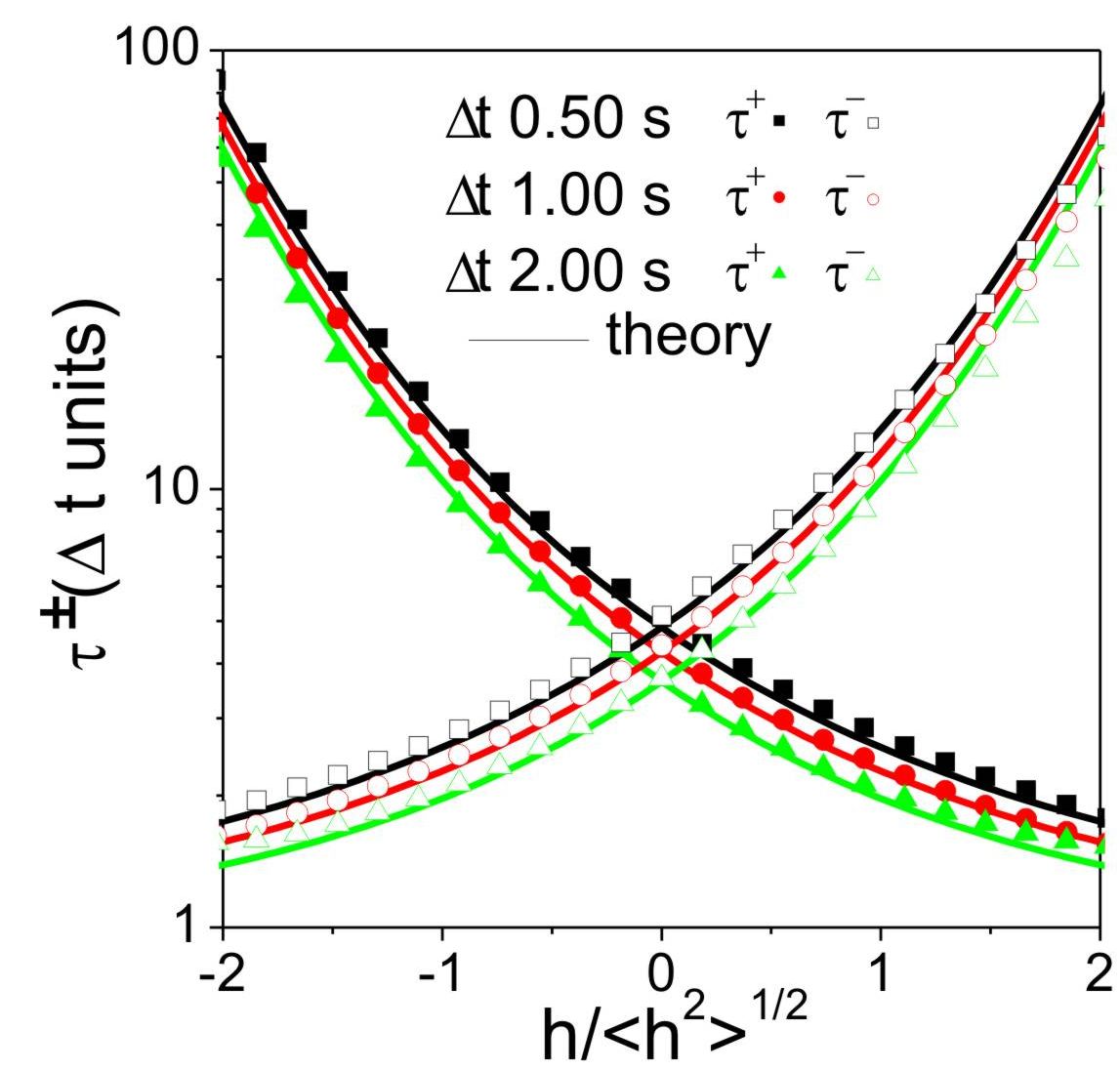}
    \caption{values of $\tau^\pm (h)$ as function of $h$ for three time interval for
three interval $\Delta t, 2 \Delta t$ and $4 \Delta t$.}
\label{taus} 
\end{center}
\end{figure}

Finally we plot in Figure \ref{taus} the dependence of the $\tau$'s on $h$ for
three choices of the time interval. The curves are confusing at first sight, since
the values of $\tau^{\pm} (h)$ are about the same for all three choices. So, if we
multiply them with the value of the chosen time interval, in order to convert
them from numbers to real times, we get substantial different times. This 
indicates that the residence and waiting time depend strongly on the measuring
process.  

The fact that the smallest chosen time interval leads to the smallest value 
of the residence and waiting time, naturally poses the question what will 
happen in the limit of vanishingly small time interval $\Delta t$. 
Theoretically it relies on the behavior of the correlation function
$g(t)$ in the limit $t \rightarrow 0$. We presented in (\ref{c14}) the 
behavior as it follows from capillary wave theory. A linear approach of $g$ to 1
implies that $\tau^{+-} (h)$ increases as the inverse power of the square root 
of $\Delta t$. Then, after multiplying with $\Delta t$ in order to get their 
values in real time, the residence and waiting time vanish as the square root 
of $\Delta t$. However, the slope of the linear term in (\ref{c14}) depends on
a molecular quantity $\kappa$, which indicates that wavelengths matter
for which the mesoscopic capillary waves theory is not designed \cite{weeks}. 
One could argue that for molecular times, the cusp in 
$g(0,t)$ is rounded off to a parabola (since it is time reversal invariant).
Then this parabola would compensate the square root in (\ref{f4}) 
and the residence and waiting time would approach a finite limit. 

Unfortunately this scenario can not be tested experimentally, given the 
present limits on the  sampling frequency. However, there is an interesting 
sampling regime beyond our data, for which the capillary wave theory still holds.
In Fig. \ref{mounnum} the data go up to the value $g \simeq 0.8$. The typical
square root decay of the curves for $q^{+-} (h,g)$ cannot be tested with
our data. A microscope, which is faster by a factor 10, could enter this 
regime where the typical signature of the Brownian character of 
the fluctuations is most significant. They give increasingly larger weight to 
short living hills and valleys, which force the mean values to shrink in a 
specific way predicted by the presented theory.

The same issue presents itself in the analysis of the data for a single time
as function of the sampling distance, but in a less severe way. In the appendix 
it is shown that the height-height correlation as function of the distance is 
a parabola for short distances. Thus a finite value of the sequence length
and the recurrence length would follow in the limit of continuous sampling.
However, again we do not reach the regime and the theoretical limiting
values are strongly dependent on the cut-off $\kappa$, where the capillary
wave theory breaks down.

\section{Dependence of $p^{\pm}_n (h)$ on $n$ and $h$}\label{distri}

\begin{figure} 
\begin{center}
    \centering
    \includegraphics[width=12 cm]{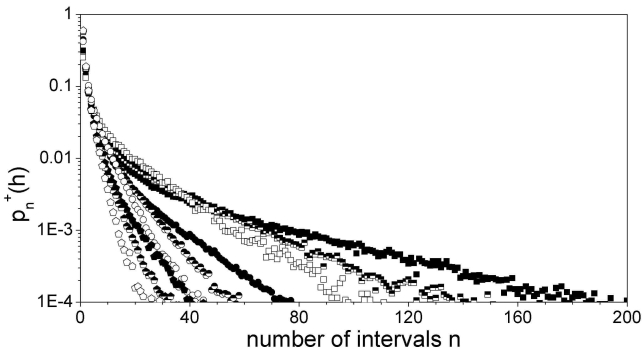}
    \caption{The temporal $p^+_n (h)$ for statepoint IV, for $h=-1$ (filled
squares), $h=0$ (circles) and $h=1$ (semifilled pentagons). The colours 
correspond to time intervals $\Delta t =$ 0.5 seconds (black symbols), 
$2 \Delta t$ (red symbols) and $4 \Delta t $ (green symbols)}
\label{distribution} 
\end{center}
\end{figure}

We plot in Fig. \ref{distribution} the experimental curves for $p^+_n (h)$
for a large number of $n$ for statepoint IV. A noteworthy point is that only
for rather large values of $n$ the decay with $n$ (time) becomes exponential. 
The scatter in the data is modest, even for large $n$ corresponding to large 
times $t$. Thus the experiment provides a host of detailed
information on the statistics of the fluctuations in a wide time range. 

\begin{figure}
\begin{center}
    \centering
    \includegraphics[width=12 cm]{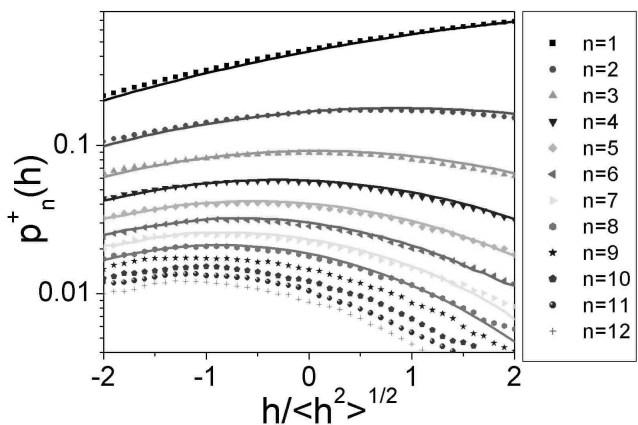}
    \caption{The temporal $p^+_n (h)$ as function of $h$ for a number of $n$ for statepoint 
IV. The drawn lines are the theoretical values.}
\label{pnplut} 
\end{center}
\end{figure}

Another way of plotting the data is to select one value of $n$ and plot 
$p^\pm _n (h) $ as function of $h$. Fig. \ref{pnplut} shows the experimental 
data for statepoint IV for $p^+_n (h)$.
This way of presenting the data facilitates the comparison with the theory.
The drawn lines are the theoretical values as given by (\ref{e8}). 
We reiterate that the input in the calculations is a set
of $n+2$ experimental values of $g(0,t)$. The agreement between theory and 
experiment is remarkable for these detailed data. Statepoint II gives 
similar results with a slight asymmetry between up and down, of the same
type as deviations in the spatial correlation functions, shown in Fig. \ref{pnplu}.

The data for $p^-_n (h)$ have been independently collected. The symmetry 
(\ref{e6}) is very well obeyed, such that there is no point in showing these 
data separately. 

\section{Discussion} \label{discus}

The above given analysis of the statistics of interface fluctuations 
naturally falls into two parts, in which the height-height correlation function 
$g(x,t)$ plays a pivotal role. The first part concerns the connection between 
$g(x,t)$ and the state parameters such as $\Delta \rho, \eta$ and $\gamma$. 
The second part is the determination of the multiple correlation-functions 
$G(h_1,\cdots , t_n)$ from $g(x,t)$ through (\ref{d4}). In the first part 
we have used the data for $g(x,t)$ to determine the state parameters.
Even though the derivation of the structure  of $g(x,t)$ in space and in time is
quite different, the behavior is remarkably
similar if the space and time variables are properly scaled
(see (\ref{cx}) and (\ref{c12})).
  
The second part has been our main concern. We used
the measured $g(x,t)$ as input, providing
all the necessary information on the statistics
of the snapshots. The advantage of splitting the problem into these two
parts is that the second part is not confounded by errors in the first. 
The only assumption in the theory is the use of the Langevin equation 
for the effect of the thermal (white) noise. The best justification for this 
procedure is {\it a posteriori} through its consequences. 
In view of the successful agreement with the experimental
results, the assumption appears to be very well fulfilled.

Experimentally the capillary waves are disentangled from the structure of
the intrinsic interface. Most amazing is that detailed correlations
in capillary waves can be determined with high accuracy.
The statistics of the temporal dependence is generally better than that of
the spatial behavior. 
We have chosen only a limited set of obtainable correlation functions in order
to compare them with the theoretical calculations. Experimentally it is easy to 
collect data for practically any interesting $n$. In Fig. \ref{pnplu} and 
Fig. \ref{distribution}  we show the distribution $p^+_n (h)$ as function 
of $n$ for the values $h=-1,0$ and 1. 

There is a simplifying aspect in the fact that experimentally only sequences
of finite time intervals can be measured. So one does not know what the
interface does in between two snapshots. But this is precisely the reason why 
it suffices to calculate the correlation functions defined in Section 
\ref{sequence}. Here also one does not specify the evolution in between two 
snapshots. For instance, the key quantity $q^{+-} (h,g)$ for the residence and 
waiting times, involves the crossing of the level $h$ by the interface. 
But it does not say that it may cross it only once! 
Any odd number of crossings is possible. In Fig. \ref{mounnum}, where
we compare $q^{+-} (h,g)$ with experiment, large time intervals feature 
(corresponding to small $g$) and in these large time intervals crossings are 
frequently taking place. Also the hills and valleys of length $n$ for which 
the distribution is given in Fig. \ref{pnplu} and Fig. \ref{pnplut}, may be 
interrupted by opposite values in between snapshots. The charm of the
comparison is that both theory and experiment allow these possibilities. 

In this paper we have restricted ourselves to sequences of height measurements
at the same time or at the same position. The general result (\ref{d4}) shows 
that one could equally well correlate snapshots at combinations of times and 
positions and do a similar statistical analysis. 
The only point that matters is the height-height correlation $g_{i,j}$ 
between the events $i$ and $j$. Also one does not need to worry in how 
much the measurements refer to a single point or to an area of finite size. 
These more collective variables are also linear combinations of the basic 
variables $h_{\bf k}$ and therefore also gaussian randomly
distributed. Then taking the measured correlations between the more general
variables as input, leads to exactly the same analysis as given here for point 
variables.

{\bf Acknowledgments}
The authors are indebted for valuable discussions with D.G.A.L. Aarts, 
H.W.J. Bl\"ote, G.T. Barkema and members of the ``Theorieclub'', where this 
problem arose. We also thank J. de Folter, C. Vonk, S. Sacanna and B. Kuipers 
for help with the synthesis and characterisation of the experimental system.
The work of VWAdV is part of the research program of the 'Stichting voor 
Fundamenteel Onderzoek der Materie (FOM)', which is financially supported by 
the 'Nederlandse Organisatie voor Wetenschappelijk Onderzoek (NWO)'.
Support of VWAdV by the DFG through the SFB TR6 is acknowledged.

\appendix

\section{The height-height correlation integral}

In this appendix we discuss some properties of the height-height correlation
function $g(x,t)$. We start with the equal time function $g(x,0)$. Using 
the scaled integration variable $y = k \xi$ the integral for $g(x,0)$ leads to
\begin{equation} \label{B1}
g(x,0) = {2 \over \log(1+\kappa^2)}
\int^{\kappa}_0 y dy {J_0 (xy/\xi) \over 1 + y^2} 
\end{equation} 
where $\kappa = k_{max} \xi$ is the cut-off. Sending this value to $\infty$
yields the modified Besselfunction $K_0$
\begin{equation} \label{B2}
\int^{\infty}_0 y dy {J_0 (xy/\xi) \over 1 + y^2} = K_0 (x/\xi)
\end{equation} 
For finite $\kappa$ we can make a short distance expansion, reading
\begin{multline} \label{B3}
\int^\kappa_0 y dy {J_0 (xy/\xi) \over 1 + y^2}= {1 \over 2} 
\log(1+\kappa^2) + \\
\left({ x \over \xi}\right)^2 
\left({\kappa^2 - \log(1+\kappa^2) \over 2 }\right)+ \cdots
\end{multline}
Matching the small argument expansion of $K_0 (x/ \xi)$
\begin{equation} \label{B4}
K_0 (x /\xi) = \log (x/2 \xi) -0.57721 + \cdots
\end{equation}   
with (\ref{B3}) gives for the point where they cross the approximate value
\begin{equation} \label{B5}
x   \simeq  {\xi \over \kappa} 
\end{equation} 

The behavior in the time direction is remarkably similar to the spatial 
direction, although the integral for $g(0,t)$ looks quite different. We write
\begin{equation} \label{A1}
g(0,t)= {2 \over \log(1+\kappa^2)} H(t/t_c ; \kappa),
\end{equation} 
with $H(z ; \kappa)$ as the  integral
\begin{equation} \label{A2}
H(z ; \kappa) = \int^{\kappa}_0 
{y d y \over 1 + y^2} \exp [- z (y + y^{-1} )/2].
\end{equation}
The first point is to prove relation (\ref{c12}), which we do by showing that
\begin{equation} \label{A3}
{d H (z, \infty) \over dz} = - K_1 (z).
\end{equation} 
and checking the asymptotic expansion of (\ref{c12}) for large $z$.
The advantage of (\ref{A3}) is that 
\begin{equation} \label{A4}
{d H (z ; \infty) \over d z}= -{1 \over 2} \int^\infty_0 dy 
\exp [- z (y + y^{-1}) /2 ]
\end{equation} 
is a simpler integral than (\ref{A2}). Then take $x = \ln y$ as 
integration variable, which turns (\ref{A4}) into
\begin{equation} \label{A5}
{d H(z ; \infty) \over d z}= -{1 \over 2} \int^\infty_{-\infty} d x \, 
e^x \, \exp (- z \cosh x).
\end{equation} 
Splitting the integral into pieces from $- \infty$ to 0 and from 0 to $\infty$
and changing in the first part from $x$ to $-x$, yields the relation
\begin{equation} \label{A6}
{d H(z; \infty) \over d z} = - \int^\infty_0 d x \cosh x 
\exp (- z \cosh x).
\end{equation} 
The integral is a representation of the function $K_1 (z)$ \cite{abramowitz}.

In order to show that no constant is lost in going from (\ref{c12}) to 
(\ref{A3}) one can check the asymptotic expansion of (\ref{c12}) for large 
$z$, which follows from an expansion around the slowest mode for $y=1$:
\begin{equation} \label{A7}
y + y^{-1} = 2 + (y -1 )^2 + \cdots 
\end{equation} 
and replacing the integral by a full gaussian around $y = 1$. Then one gets
\begin{multline} \label{A8}
H(z ; \infty) \simeq {e^{-z} \over 2} \int^\infty_{-\infty} d (y -1) \\
\exp [-(y -1)^2 z /2] = e^{-z} \left({\pi \over 2 z}\right)^{1/2},
\end{multline} 
which matches the asymptotic expansion of $K_0 (z)$.

The final point is the expansion for small times $t/t_c$ for a finite value
of $\kappa$. We expand the exponential
\begin{equation} \label{A9}
\exp [ -(y + y^{-1}) z /2] = 1 -(y + y^{-1}) z /2 + \cdots
\end{equation} 
and insert this expansion into the integral (\ref{A2}). Then we find for $H(z ; \kappa)$
\begin{equation} \label{A10}
H(z ; \kappa) = {1 \over 2} \ln(1 + \kappa^2) - {\kappa \over 2} \, z 
+ \cdots
\end{equation} 
Note that the next term in this expansion leads to a logarithmically divergent 
integral at the small $y$ side. Thus the next term is not of the order $z^2$
but of the order $z^2 \ln z$. The finite $\kappa$ integral stays
finite in contrast to $K_0 (z)$ which diverges for $z=0$. 
With the expansion (\ref{B4}) and (\ref{A10}) we find for the point where the 
finite-$\kappa$ curve starts to deviate from the $K_0 (z)$:
\begin{equation} \label{A11}
t \simeq t_c /\kappa.
\end{equation} 
One obtains the rough estimate $z \simeq 1 /\kappa $ for this matching point, 
by looking to the value of the exponential at the upper
boundary, which is $\exp [- z (\kappa + \kappa^{-1})/2]$. For larger values the 
boundary value starts to vanish and extending the integral to infinity leads
to a small error. For smaller values of this $z$, the exponent of the exponential 
becomes smaller than 1 at the boundary and the integrand of (\ref{A2}) has not yet 
died out at $y = \kappa$. Then deviations from the infinite domain start to show up.

\end{document}